\documentclass[intlimits,twoside,a4paper]{article}

\usepackage{amsmath,amssymb}
\usepackage{graphicx}
\usepackage[T2A]{fontenc}
\usepackage[cp1251]{inputenc}

\usepackage{cmpj2}

\issue{2015}{18}{3}{33704}
\doinumber{10.5488/CMP.18.33704}

\title[Effects of correlated hopping on electronic ferroelectricity]%
{Effects of correlated hopping on electronic ferroelectricity
in the extended Falicov-Kimball model in two dimensions}
\author[P. Farka\v sovsk\'y, J. Jure\v ckov\'a]%
{P. Farka\v sovsk\'y\thanks{E-mail: farky@saske.sk}\,, J. Jure\v ckov\'a}

\address{Institute of Experimental Physics, Slovak Academy of Sciences, Watsonova 47, 040 01 Ko\v {s}ice, Slovakia}

\authorcopyright{P. Farka\v sovsk\'y, J. Jure\v ckov\'a, 2015}
\date{Received April 29, 2015}

\begin{document}

\maketitle

\begin{abstract}
We use the Hartree-Fock (HF) approximation with the charge-density-wave (CDW)
instability to study the effect of correlated hopping on the stability
of electronic ferroelectricity in the extended Falicov-Kimball model (FKM)
 in two dimensions. It is shown that the effects of correlated
hopping term are very strong, especially for negative values of the
$f$-level energy $E_f$, where they lead to: (i) stabilization of the
ferroelectric ground state with a spontaneous hybridization $P_{df}=\langle
d^+f \rangle$ for positive values of correlated hopping parameter $t'_d$,
(ii) stabilization of the antiferroelectric phase for
$t'_d < t'_{d,\textrm{crit.}} < 0$ and (iii) suppression of the ferroelectric
ground state for $t'_{d,\textrm{crit.}}  \leqslant t'_d < 0$, The effects of correlated
hopping on valence transitions are also discussed.

\keywords electronic ferroelectricity, charge density waves, valence transitions

\pacs 71.27.+a, 71.28.+d, 71.30.+h
\end{abstract}
 	
 \vspace{5mm}
\section{Introduction}

In the recent years the Falicov-Kimball model (FKM)~\cite{Falicov} has been
extensively studied in connection with the exciting idea of electronic
ferroelectricity~\cite{P1,P2,Cz,F1,F2,Zl,B1,B2,F3,Schneider} that is also
directly related to the problem of an excitonic
insulator~\cite{Z1,Phan,Seki,Z2,Kaneko1,Kaneko2,Apinyan}.
It is generally supposed that ferroelectricity in mixed-valent
compounds is of purely electronic origin, i.e., it results from an electronic
phase transition, in contrast to the conventional displacive
ferroelectricity
due to a lattice distortion. Since the FKM is probably the simplest model of
electronic phase transitions in rare-earth  and transition-metal compounds
it was natural to test the idea of electronic ferroelectricity just
on this model.

The FKM is based on the coexistence of two
different types of electronic states in a given material:
localized, highly correlated ionic-like states and extended,
uncorrelated, Bloch-like states. It is accepted that
insulator-metal transitions result from a change in the occupation
numbers of these electronic states, which remain themselves basically
unchanged by their character. Taking into account only the
intra-atomic Coulomb interaction between the two types of states,
the Hamiltonian of the spinless FKM can be written as the sum
of three terms:

\begin{equation}
H=\sum_{ij}t_{ij}d^+_id_j+U\sum_if^+_if_id^+_id_i+E_f\sum_if^+_if_i\,,
\end{equation}
where $f^+_i$, $f_i$ are the creation and annihilation
operators  for an electron in  the localized state at
lattice site $i$ with binding energy $E_f$, and $d^+_i$,
$d_i$ are the creation and annihilation operators
of the itinerant spinless electrons in the $d$-band
Wannier state at site $i$.

The first term of (1) is the kinetic energy corresponding to
quantum-mechanical hopping of the itinerant $d$ electrons
between sites $i$ and $j$. These intersite hopping
transitions are described by the matrix  elements $t_{ij}$,
which are $-t_d$ if $i$ and $j$ are the nearest neighbors and
zero otherwise (in what follows all parameters are measured
in units of $t_d$). The second term represents the on-site
Coulomb interaction between the $d$-band electrons with density
$n_d=\frac{1}{L}\sum_id^+_id_i$ and the localized
$f$ electrons with density $n_f=\frac{1}{L}\sum_if^+_if_i$,
where $L$ is the number of lattice sites. The third  term stands
for the localized $f$ electrons whose sharp energy level is $E_f$.

The first attempt to describe the electronic ferroelectricity
within the FKM was made by Portengen~et~al.~\cite{P1,P2}.
They studied the FKM with a $k$-dependent
hybridization in Hartree-Fock (HF) approximation and found, in particular,
that the Coulomb interaction $U$ between the itinerant $d$-electrons
and the localized $f$-electrons gives rise to a non-vanishing excitonic
$\langle d^{+}f \rangle $-expectation value even in the limit of vanishing
hybridization $V \rightarrow 0$. As an applied (optical) electrical field
provides for excitations between $d$- and $f$-states and thus for an excitonic
expectation value $P_{df}=\langle d_i^{+}f_i \rangle$, the finding of a
spontaneous $P_{df}$ (without hybridization or electric field) has been
interpreted as the evidence for electronic ferroelectricity.
However, analytical calculations within well controlled approximation (for
$U$ small) performed by Czycholl~\cite{Cz} in infinite dimensions do
not confirm the existence of electronic ferroelectricity.
The same conclusion has been also
obtained independently~\cite{F1,F2} by small-cluster exact-diagonalization
and density-matrix renormalization group calculations in the one dimension
for both small ($U<1$) and intermediate interactions ($U\sim5$).

Hybridization between the itinerant $d$ and localized $f$ states, however,
is not the only way to develop $d$--$f$ coherence. Recent theoretical works
by Batista et al.~\cite{B1,B2} showed that the ground state with
a spontaneous electric polarization can also be induced by the
nearest-neighbor $f$-electron hopping ($-t_f\sum_{\langle i,j\rangle}f^+_if_j$)
for dimensions $D>1$.
Based on these results the authors postulated
the following conditions that favor the formation of the electronically
driven ferroelectric state: (i) the system should be in a mixed-valence
regime, (ii)  two bands involved should be of different parity, (iii) a
local Coulomb repulsion $U$ between the $f$ and $d$ orbitals is required.

Later on this model was extensively used to describe different phases
in the ground state and particularly the properties of the excitonic
phase~\cite{Z1,Phan,Seki,Z2,Kaneko1,Kaneko2,Apinyan,Ejima}.
It was found that the ground state phase diagram
exhibits a very simple structure consisting of only four phases, and
namely, the full $d$ and $f$ band insulator (BI), the excitonic
insulator (EI), the charge-density-wave (CDW) and the  staggered
orbital order (SOO). The EI is characterized by a nonvanishing
$\langle d^+f \rangle$ average. The CDW is described by a periodic
modulation in the total electron density of both $f$ and $d$ electrons,
and the SOO is characterized by a periodic modulation in the
difference between the $f$ and $d$ electron densities.

In our recent paper~\cite{F3}, the HF approach with the CDW instability
was used to study the ground-state properties of the extended FKM with
$f$--$f$ hopping in two and three dimensions. It was found that the HF
solutions with the CDW instability perfectly reproduce the two-dimensional
intermediate coupling phase diagram of the model calculated  by constrained
path Monte Carlo (CPMC) method~\cite{B2}, including phase boundaries
of ferroelectric state. On the other hand, it should be noted that the model
discussed above neglects all nonlocal interaction terms, and thus it
is questionable whether the above mentioned results also persist in
more realistic situations when nonlocal interactions are turned
on. An important nonlocal interaction term obviously absent
in the conventional FKM is the term of correlated hopping,
\begin{equation}
H=-t'_d\sum_{\langle ij\rangle}d^+_id_j(f^+_if_i+f^+_jf_j),
\end{equation}
in which the $d$-electron hopping amplitudes between the neighboring
lattice sites $i$ and $j$ explicitly depend on the occupancy $(f^+_if_i)$
of the $f$-electron orbitals.
The importance of the correlated hopping term has been already
mentioned by Hubbard~\cite{Hubbard}. Later Hirsch~\cite{Hirsch} pointed
out that this term may be relevant in explaning the superconducting properties of strongly correlated electrons.
Here, we examine the effects of this term on the stability of the EI phase
in the  two-dimensional extended FKM. With the exception of small deviations,
the method used bellow is the same as the one described in our previous
paper~\cite{F3} and, therefore, we summarize here only its main steps.

\clearpage

\section{The method}

The Hamiltonian of the extended FKM with correlated hopping has the form
\begin{equation}
H=-t_d\sum_{\langle ij \rangle }d^+_id_j-t_f\sum_{\langle ij \rangle }f^+_if_j+U\sum_if^+_if_id^+_id_i
+E_f\sum_if^+_if_i-t'_d\sum_{\langle ij \rangle }d^+_id_j\left(f^+_if_i+f^+_jf_j\right).
\end{equation}
In accordance with our previous paper~\cite{F3},
we go here beyond the usual HF approach~\cite{Leder} in which
only homogeneous solutions are postulated and also consider inhomogeneous
solutions modelled by a periodic modulation of the order parameters:
\begin{equation}
\langle n_i^f\rangle=n_f+\delta_f\cos({\bf Q}\cdot {\bf r}_i),\quad
\langle n_i^d\rangle=n_d+\delta_d\cos({\bf Q}\cdot {\bf r}_i),\quad
\langle f_i^+d_i\rangle=\Delta+\Delta_P\cos({\bf Q}\cdot {\bf r}_i).
\end{equation}
Here, $\delta_{d}$ and $\delta_{f}$ are the order parameters of the CDW state
for the $d$- and $f$-electrons, $\Delta$ is the excitonic average and
${\bf{Q}}=(\pi,\pi)$ is the nesting vector for $D=2$.

Using these expressions, the HF Hamiltonian of the extended
FKM with correlated hopping can be written as
\begin{eqnarray}
{\cal H}&=&(-t_d-2t'_dn_f)\sum_{\langle i,j\rangle} d^+_id_j - t_f\sum_{\langle i,j\rangle}f^+_if_j
+ E_f\sum_i n_i^f
      + U\sum_i \left[n_f+\delta_f\cos({\bf Q}\cdot {\bf r}_i)\right]n_i^d
\nonumber\\
      &&+ \, U\sum_i \left[n_d+\delta_d\cos({\bf Q}\cdot {\bf r}_i)\right]n_i^f
      -U\sum_{i} \left[ \Delta+\Delta_P\cos({\bf Q}\cdot {\bf r}_i)\right]
      d_i^+f_i + \text{h.c.}\,,
\end{eqnarray}
where the simple decoupling of the form
\begin{equation}
d^+_id_j(f^+_if_i+f^+_jf_j) \to d^+_id_j(\langle
f^+_if_i\rangle+\langle f^+_jf_j\rangle)=2n_fd^+_id_j
\end{equation}
has been used for the correlated hopping term.
Thus, one can see that within this decoupling scheme, the effect of the correlated
hopping is fully transformed to the renormalization of the static single
electron hopping amplitude according to $t_d \to t^*_d =
t_d+2t'_d n_f$. Next, we shall show that this relatively small change
has dramatical effects on the existence of the EI phase
in the extended FKM.

This Hamiltonian can be  diagonalized by the following
canonical transformation~\cite{F3,Brydon}
\begin{eqnarray}
\begin{array}{ccc}
\gamma_k^m=u_k^md_k + v_k^md_{k+{Q}} + a_k^mf_k + b_k^mf_{k+{Q}}\, , &   &
m=1,2,3,4\, ,
\end{array}
\end{eqnarray}
where $\Psi_k^m=(a_k^m,b_k^m,u_k^m,v_k^m)^T$ are solutions of the associated Bogoliubov-de Gennes (BdG)
eigenequations:

\begin{equation}
H_k\Psi_k^m=E_k^m\Psi_k^m\, ,
\end{equation}
with
\begin{eqnarray}
H_k=\left( \begin{array}{cccc}
    \epsilon_k^d+Un_f &  U\delta_f &   -U\Delta &     -U\Delta_P \\
     U\delta_f  & \epsilon_{k+Q}^d+Un_f & -U\Delta_P &    -U\Delta\\
     -U\Delta   &  -U\Delta_P   &  \epsilon_{k}^f+Un_d+E_f & U\delta_d \\
     -U\Delta_P &  -U\Delta    &  U\delta_d  & \epsilon_{k+Q}^f+Un_d+E_f\\
            \end{array}
\right),
\end{eqnarray}
and the corresponding energy dispersions $\epsilon_k^d$ and $\epsilon_k^f$
for  $d$ and $f$ electrons and the HF parameters $n_d$, $\delta_d$, $n_f$, $\delta_f$, $\Delta$, $\Delta_P$ are given by:
\begin{equation}
\epsilon_k^{f}=-2t_f\left[\cos(k_x)+\cos(k_y)\right],\qquad \epsilon_k^{d}=-2t^*_d\left[\cos(k_x)+\cos(k_y)\right],
\end{equation}
\begin{equation}
n_d=\frac{1}{N}\sum_k{}'\sum_m \{ u_k^mu_k^m + v_k^mv_k^m\}f(E_k^m),\qquad
\delta_d=\frac{1}{N}\sum_k{}'\sum_m \{ v_k^mu_k^m + u_k^mv_k^m\}f(E_k^m),
\end{equation}
\begin{equation}
n_f=\frac{1}{N}\sum_k{}'\sum_m \{ a_k^ma_k^m + b_k^mb_k^m\}f(E_k^m),\qquad
\delta_f=\frac{1}{N}\sum_k{}'\sum_m \{ b_k^ma_k^m + a_k^mb_k^m\}f(E_k^m),
\end{equation}
\begin{equation}
\Delta=\frac{1}{N}\sum_k{}'\sum_m \{ a_k^mu_k^m + b_k^mv_k^m\}f(E_k^m),\qquad
\Delta_P=\frac{1}{N}\sum_k{}'\sum_m \{ b_k^mu_k^m + a_k^mv_k^m\}f(E_k^m).
\end{equation}
Here, the  prime denotes summation over half the Brillouin zone and
$f(E)=1/\{1+\exp[\beta(E-\mu)]\}$ is the Fermi distribution function.


\section{Results and discussion}

To examine the effects of correlated hopping on the ground-state phase diagram of the extended FKM
(in the $E_f$--$t_f$ plane), and particularly on the stability of the electronic
ferroelectricity, we use the zero temperature variant of the method described
above. The HF equations are solved self-consistently for each pair of $(E_f,t_f)$ values
for several selected values of $t'_d$. We use an exact diagonalization method to solve
the Bogoliubov-de Gennes equation. We start with an initial set of the order parameters.
By solving equation (8), the new order parameters are computed via equations (11) to
(13) and are substituted back into equation (8). The iteration is repeated until a desired accuracy
is achieved. It should be noted that the stability of different HF solutions
has been also checked numerically by calculating the total energy and it was
found that all phases presented in the ground-state phase diagram represent
the most stable HF solutions.

\begin{figure}[!b]
\begin{center}
\includegraphics[width=9.5cm]{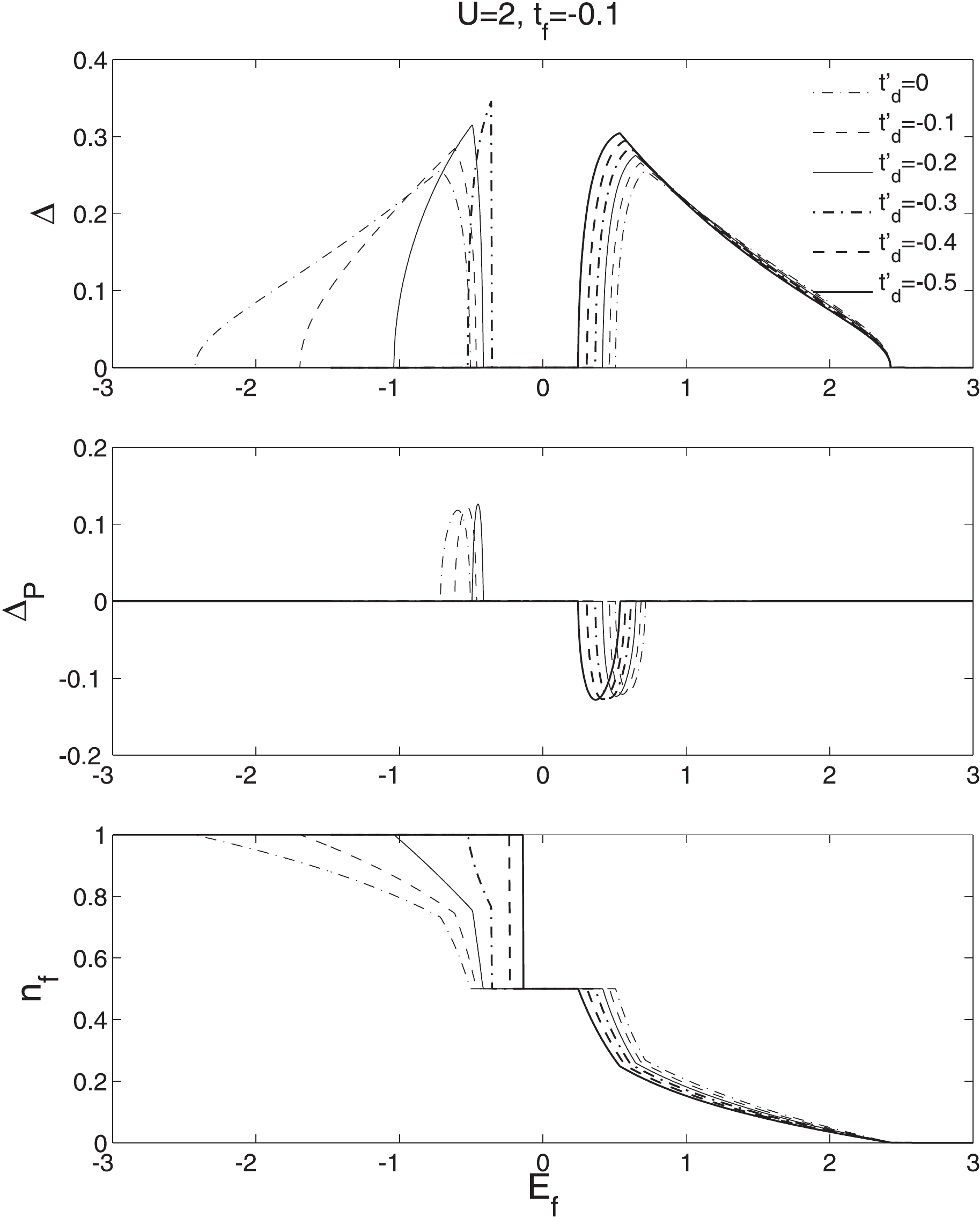}
\end{center}
\vspace*{-0.4cm}
\caption{\looseness=-1Dependence of the HF parameters $\Delta$, $\Delta_P$ and
$n_f$ on the $f$-level energy $E_f$ calculated for different values of $t'_d$
($t'_d=0$, $-0.1$, $-0.2$, $-0.3$, $-0.4$, $-0.5$) and $U=2$, $t_f=-0.1$.}
\label{fig1}
\end{figure}

First, we have examined the model in the intermediate coupling regime $U=2$,
since for this case, there exists a comprehensive phase diagram of the
extended FKM model ($t_f \neq 0, t'_d=0$) obtained by a CPMC technique~\cite{B2}
as well as its HF version \cite{F3} and both accord very nicely (both qualitatively and quantitatively). In figure~\ref{fig1}, we have displayed typical examples of our HF solutions
obtained for $\Delta$, $\Delta_P$  and $n_f$ in the limit
of small $t_f$ values ($t_f=-0.1$) and $t'_d<0$.
One can see that going with $t'_d$
from 0 to $t'_{d,\textrm{crit.}}=-0.5$, the effects of correlated hopping on the stability
of excitonic phase ($\Delta \neq 0$) increase dramatically, especially
in the limit of $E_f <0 $. For negative $E_f$, the stability region of excitonic
phase rapidly reduces with increasing $|t'_d|$, and at $t'_d \sim -0.4$, it
fully disappears. This is obviously a consequence
of the absence of a mixed valence phase for $n_f > 0.5$ and $t'_d \leqslant -0.4$
as it is demonstrated in Fig~1, where the average $f$-electron occupancy
$n_f$ is plotted as a function of the $f$-level energy $E_f$.
In this case, the valence transitions change discontinuously from
$n_f=1$ to $n_f=0.5$
and the system exhibits a direct transition from the BI phase to the CDW
phase. At the same time,  this result stresses the fact what a crucial role is played by
the term of correlated hopping in the mechanism of valence transitions,
and that already very small values of the correlated hopping parameter
are capable of changing the type of valence transitions from continuous
to discontinuous. For this reason, the term of correlated hopping
should be surely taken into account in the correct
description of valence transitions in rare-earth
compounds.

\begin{figure}[!h]
\begin{center}
\includegraphics[width=9.5cm]{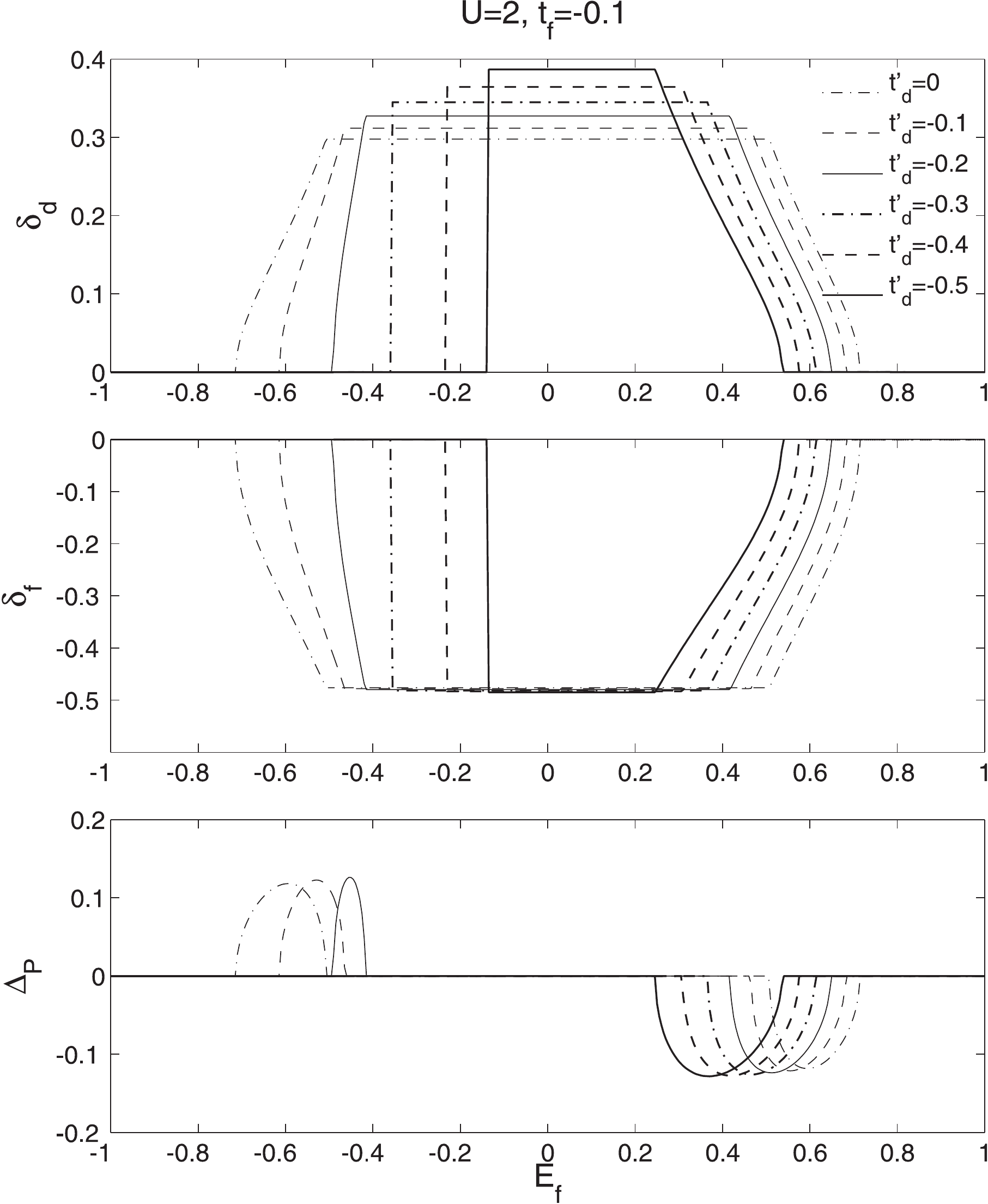}
\end{center}
\vspace*{-0.4cm}
\caption{Dependence of the HF parameters $\delta_d$, $\delta_f$ and
$\Delta_P$ on the $f$-level energy $E_f$ calculated for different values of $t'_d$
($t'_d=0$, $-0.1$, $-0.2$, $-0.3$, $-0.4$, $-0.5$) and $U=2$, $t_f=-0.1$.}
\label{fig2}
\end{figure}

In accordance with the case $t'_d = 0$, we have found two different
excitonic phases for $t'_d < 0$. The first one is homogeneous
($\Delta > 0,\Delta_P=0$) and the second one is inhomogeneous
( $\Delta > 0,\Delta_P \neq 0$).
Analysing the numerical data obtained for
$\delta_d, \delta_f$ and $\Delta_P$ (see figure~\ref{fig2}), one can see that the
appearance of an inhomogeneous phase is tightly connected with the formation of the
CDW ordering in $f$ and $d$ electron subsystems since in the corresponding
regions where $\Delta_P \neq 0$, the order parameter $\delta_d$ ($\delta_f$)
changes continuously from 0 to its maximal (minimal) value. With increasing
$|t'_d|$, the stability region of the inhomogeneous phase strongly
reduces and at $t'_d \sim -0.3$ this phase practically disappears.

\begin{figure}[!t]
\begin{center}
\includegraphics[width=9.5cm]{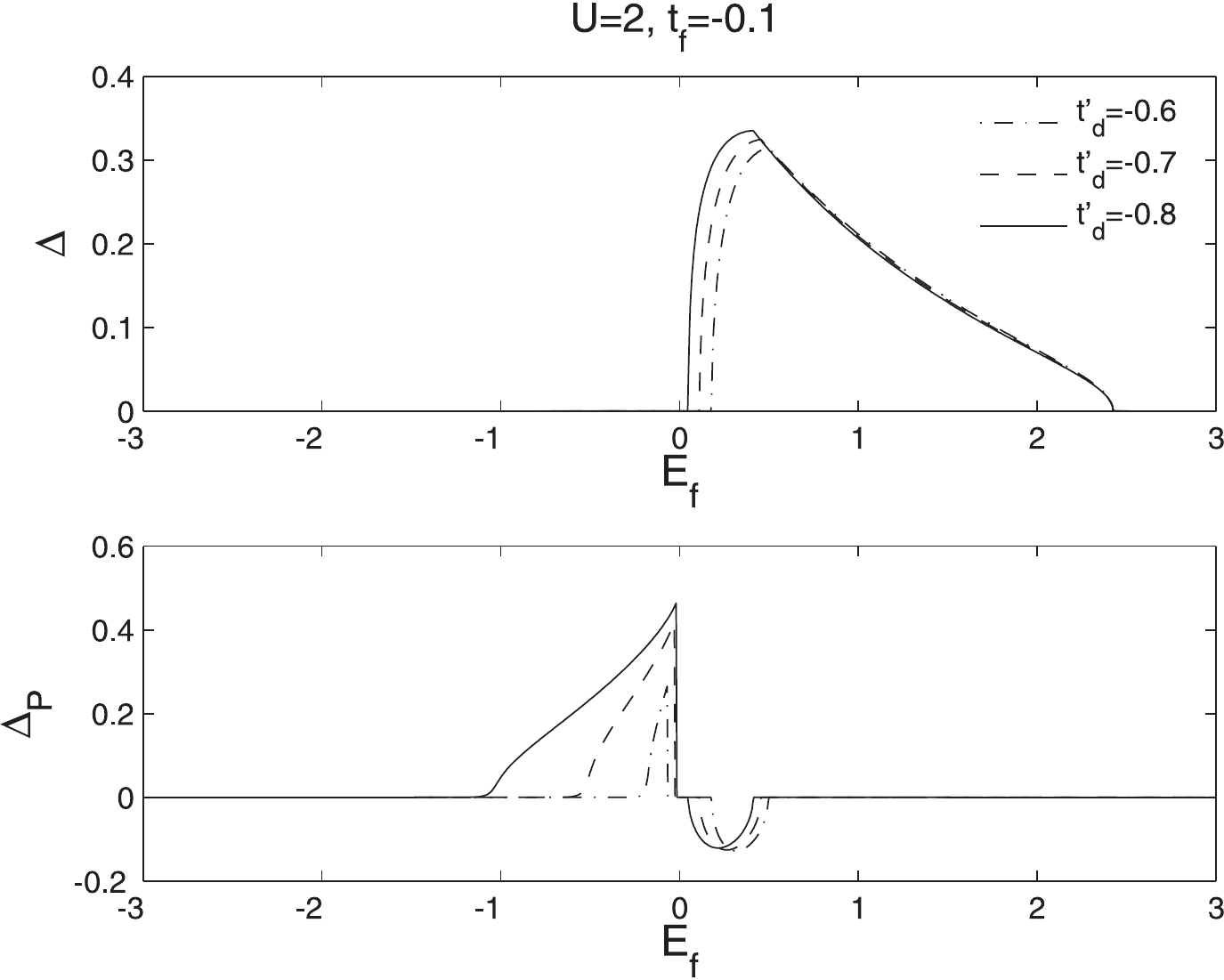}
\end{center}
\vspace*{-0.4cm}
\caption{Dependence of the HF parameters $\Delta$ and $\Delta_P$
on the $f$-level energy $E_f$ calculated for different values of $t'_d$
($t'_d= -0.6$, $-0.7$, $-0.8$) and $U=2$, $t_f=-0.1$.}
\label{fig3}
\end{figure}

\begin{figure}[!b]
\begin{center}
\includegraphics[width=9.5cm]{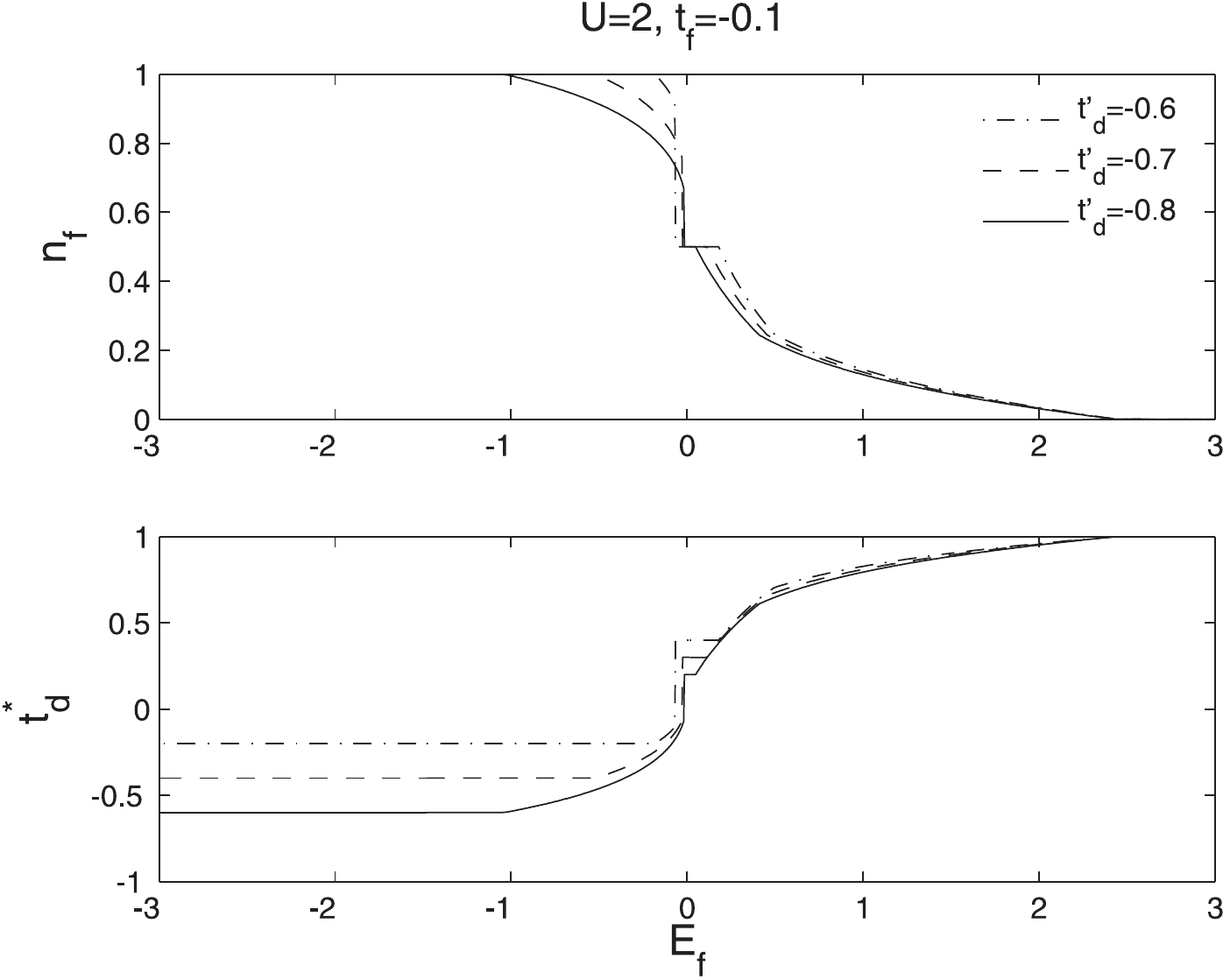}
\end{center}
\vspace*{-0.4cm}
\caption{Dependence of the HF parameter $n_f$ and the renormalized
$d$-electron hopping parameter $t^*_d$ on the $f$-level energy $E_f$ calculated
for different values of $t'_d$ ($t'_d= -0.6$, $-0.7$, $-0.8$) and $U=2$, $t_f=-0.1$.}
\label{fig4}
\end{figure}

Going with $t'_d$ to smaller values we have observed a completely different behavior
of the model in the limit of $E_f<0$ (see figure~\ref{fig3}).
For all the examined values of $t'_d$, we have found that
$\Delta = 0$ and $\Delta_P > 0$, which means that the ground state of the
model is antiferroelectric in this limit. Very interesting is also the behavior
of $n_f$ as a function of $E_f$. As is shown in figure~\ref{fig4}, the mixed-valent phase
is again stabilized in the region $n_f > 1/2$ and the magnitude of the
discontinuous valence transition from $n_f > 0.5$ to $n_f=1/2$ gradually
decreases with decreasing $t'_d$ and a similar behaviour also exhibits the
width of the $n_f=1/2$ phase.
At a first glance, this result seems to be in contradiction with the
rules favour the formation of the electronically driven ferroelectric state
postulated by Batista et al. \cite{B2}. The system is in a
mixed-valence regime, the local Coulomb interaction between different
orbitals is present, but the ferroelectric state is absent. A more detailed
analysis (see figure~\ref{fig4}) of the behaviour of the renormalized $d$-electron hopping
parameter $t^*_d$ shows, however, that there in no contradiction with
conditions postulated by Batista et al., since for $t'_d < -0.5$ and $E_f <0$,
the $d$ band changes its parity. In this region, the $d$ and $f$ bands are of
the same parity ($t^*_d < 0, t_f<0$), which, in accordance with the above mentioned
rules, does not support the ferroelectric ground state.

\begin{figure}[!t]
\begin{center}
\includegraphics[width=9.5cm]{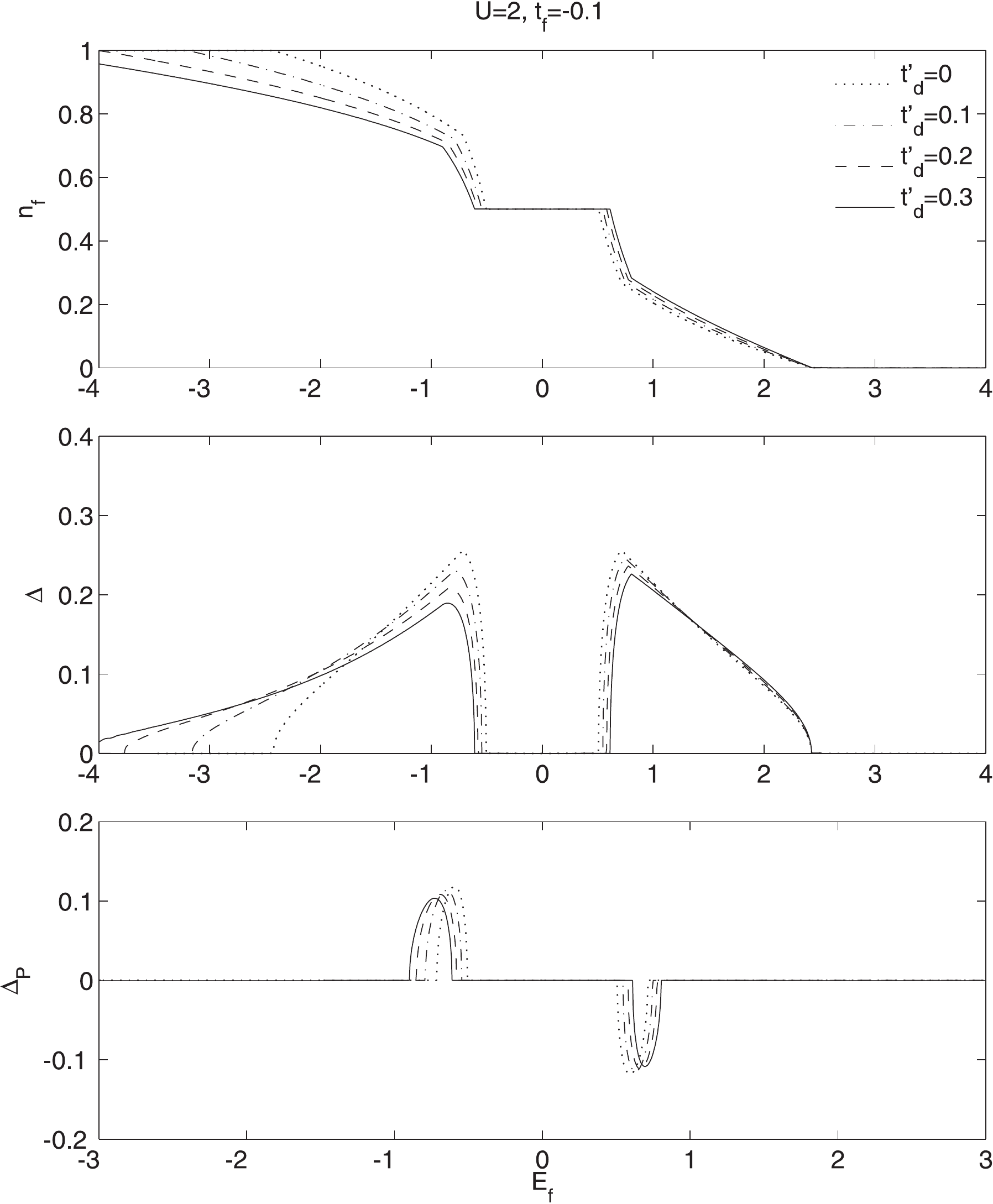}
\end{center}
\vspace*{-0.4cm}
\caption{Dependence of the HF parameters $n_f$, $\Delta$ and
$\Delta_P$ on the $f$-level energy $E_f$ calculated for different values of $t'_d$
($t'_d=0$, 0.1, 0.2, 0.3) and $U=2$, $t_f=-0.1$.}
\label{fig5}
\end{figure}

The situation in the opposite limit $t'_d > 0$ is depicted in figure~\ref{fig5}.
Again one can recognize two different regimes in the behaviour of $n_f, \Delta$
and $\Delta_p$. For positive $E_f$, only weak effects of correlated hopping
on $n_f, \Delta$ and $\Delta_p$ are observed, while in the opposite limit,
these effects are quite dramatic and the stability regions of the mixed valence
as well as excitonic phase are considerably enhanced by increasing $t'_d$.
This result independently confirms that the term of correlated hopping plays
a very important role in the  mechanism of stabilizing the excitonic phase
and, thus, it should be taken into account for the correct description of this
phenomenon. Finally, it should be noted that  we have also found qualitatively the same  behaviour
of the model for smaller values of $t_f$ and larger values
of $U$.

Thus, we can conclude that the effects of correlated hopping term on the stability
of the excitonic state in the extended FKM (within the HF
approximation with the CDW instability) are very strong and they lead to:
(i) suppression of the excitonic phase for $t'_{d,\textrm{crit}.} \leqslant t'_d < 0$,
$E_f<0$, (ii) stabilization of the antiferroelectric phase for
$t'_d < t'_{d,\textrm{crit.}}$, $E_f <0$  and (iii) stabilization of excitonic
phase for $t'_d >0$, $E_f <0$. Similarly, the strong effects of correlated
hopping are also observed on the mechanism of valence transitions, where
already very small values of the correlated hopping parameter are capable
of changing the type of valence transitions from continuous to discontinuous.

\section*{Acknowledgements}

This work was supported by the Slovak Grant Agency VEGA
under Grant No.~2/0077/13, Slovak Research and Development Agency (APVV)
under Grant APVV--0097--12 and EU Grants  No.~ITMS26210120002 and ITMS26110230097.

 \ukrainianpart

 \title{Вплив корельованого переносу на електронну
 сегнетоелектрику в розширеній моделі Фалікова-Кімбала \\ у двох
 вимірах}
 \author{П. Фаркашовскі, Ю. Юречкова}
 \address{Інститут експериментальної фізики, Словацька академія наук, 040 01
 Кошиці, Словаччина}

 \makeukrtitle

 \begin{abstract}
 \tolerance=3000%
 Ми використовуємо наближення Гартрі-Фока (ГФ) із нестійкістю щодо хвиль
 зарядової густини до дослідження впливу корельованого переносу на стійкість
 електронної сегнетоелектрики в розширеній моделі
 Фалікова-Кімбала (ФК) у двох вимірах. Показано, що внесок корельованого
 переносу є дуже сильний, особливо для від'ємних значень енергiї $f$-рiвня
 $E_f$, коли цей
 вплив приводить до: (1) стабілізації сегнето\-електричного
 основного стану зі спонтанною гібридизацією $P_{df}=\langle d^+f \rangle$ для
 додатних значень параметра корельованого переносу $t'_d$, (2) стабілізації
 антисегнетоелектричної фази при $t'_d < t'_{d,\textrm{crit}.} < 0$
 i (3) приглушення сегнетоелектричного основного стану для
 $t'_{d,\textrm{crit}.}  \leqslant t'_d < 0$. Крім того, обговорюються впливи корельованого
 переносу на переходи зі зміною валентності.
 \keywords електронна сегнетоелектрика, хвилі зарядової густини,
 переходи зі зміною валентності
\end{abstract}


\newpage

\begin{thebibliography}{99}

\bibitem{Falicov} Falicov L.M.,  Kimball J.C., Phys. Rev. Lett., 1969,
\textbf{22}, 997; \doi{10.1103/PhysRevLett.22.997}.


\bibitem{P1}  Portengen T.,  \"Ostreich T.,  Sham L.J., Phys. Rev. Lett.,
1996, \textbf{76}, 3384; \doi{10.1103/PhysRevLett.76.3384}.

\bibitem{P2}  Portengen T.,  \"Ostreich T.,  Sham L.J., Phys. Rev. B, 1996,
\textbf{54}, 17452; \doi{10.1103/PhysRevB.54.17452}.

\bibitem{Cz}  Czycholl G., Phys. Rev. B, 1999, \textbf{59}, 2642; \doi{10.1103/PhysRevB.59.2642}.

\bibitem{F1} Farka\v{s}ovsk\'y P., Phys. Rev. B, 1999, \textbf{59}, 9707; \doi{10.1103/PhysRevB.59.9707}.

\bibitem{F2}  Farka\v{s}ovsk\'y P., Phys. Rev. B, 2002, \textbf{65}, 81102; \doi{10.1103/PhysRevB.65.081102}.

\bibitem{Zl}  Zlati\'c V.,  Freericks J.K.,  Lemanski R.,  Czycholl G.,
Philos. Mag. B, 2001, \textbf{81}, 1443; \doi{10.1080/13642810110066470}.

\bibitem{B1}  Batista C.D., Phys. Rev. Lett., 2002, \textbf{89}, 166403; \doi{10.1103/PhysRevLett.89.166403}.

\bibitem{B2}  Batista C.D.,  Gubernatis J.E., Bon\v{c}a J.,  Lin H.Q.,
Phys. Rev. Lett., 2004, \textbf{92}, 187601; \\ \doi{10.1103/PhysRevLett.92.187601}.

\bibitem{F3}  Farka\v{s}ovsk\'y P., Phys. Rev. B, 2008, \textbf{77}, 155130; \doi{10.1103/PhysRevB.77.155130}.

\bibitem{Schneider}  Schneider C.,  Czycholl G., Eur. Phys. J. B, 2008, \textbf{64},
43; \doi{10.1140/epjb/e2008-00273-y}.

\bibitem{Z1}  Zenker B.,  Ihle D.,  Bronold F.X.,  Fehske H., Phys. Rev. B, 2010,
\textbf{81}, 115122; \doi{10.1103/PhysRevB.81.115122}.

\bibitem{Phan}  Phan V.N.,  Becker K.W.,  Fehske H., Phys. Rev. B, 2010,
\textbf{81}, 205117; \doi{10.1103/PhysRevB.81.205117}.

\bibitem{Seki}  Seki K.,  Eder R.,  Ohta Y., Phys. Rev. B, 2011,
\textbf{84}, 245106; \doi{10.1103/PhysRevB.84.245106}.

\bibitem{Z2}  Zenker B.,  Ihle D., Bronold F.X.,  Fehske H., Phys. Rev. B, 2012,
\textbf{85}, 121102R; \doi{10.1103/PhysRevB.85.121102}.

\bibitem{Kaneko1}  Kaneko T.,  Seki K.,  Ohta Y., Phys. Rev. B, 2012,
\textbf{85}, 165135; \doi{10.1103/PhysRevB.85.165135}.

\bibitem{Kaneko2}  Kaneko T.,  Ejima S.,  Fehske H.,  Ohta Y., Phys. Rev. B, 2013,
\textbf{88}, 035312; \doi{10.1103/PhysRevB.88.035312}.

\bibitem{Apinyan}  Apinyan V.,  Kopec T.K., J. Low Temp. Phys., 2014, \textbf{176}, 27; \doi{10.1007/s10909-014-1165-x}.

\bibitem{Ejima}  Ejima S.,  Kaneko T.,  Ohta Y.,  Fehske H., Phys. Rev. Lett., 2014,
\textbf{112}, 026401; \doi{10.1103/PhysRevLett.112.026401}.

\bibitem{Hubbard}  Hubbard J.,  Proc. R. Soc. London A, 1963, \textbf{276}, 238; \doi{10.1098/rspa.1963.0204}.

\bibitem{Hirsch}  Hirsch J.E.,  Physica C, 1989, \textbf{158}, 236; \doi{10.1016/0921-4534(89)90225-6}.

\bibitem{Leder}  Leder H.J., Solid State Commum., 1978, \textbf{27}, 579; \doi{10.1016/0038-1098(78)90399-X}.

\bibitem{Brydon}  Brydon P.M.R.,  Zhu J.X.,  Gulacsi M.,  Bishop A.R.,
Phys. Rev. B, 2005, \textbf{72}, 125122; \doi{10.1103/PhysRevB.72.125122}.

\end{thebibliography}
\end{document}